\documentstyle[preprint,revtex]{aps}

\begin{document}

\def\overlay#1#2{\setbox0=\hbox{#1}\setbox1=\hbox to \wd0{\hss #2\hss}#1%
\hskip -2\wd0\copy1}

\hfill {\bf IC/92/156}

\hfill {\bf hep-ph/9208242 }

\begin{title}
Composite Vector Leptoquarks in $e^+ e^-$, $\gamma e$, and $\gamma\gamma$
 Colliders
\end{title}

\medskip

\author{ J. E. Cieza Montalvo\cite{ci}}

\medskip

\begin{instit}
Departamento de F\'{\i}sica Matem\'atica \\
Instituto de F\'{\i}sica da Universidade de S\~ao Paulo \\
C.P. 20516, 01498 S\~ao Paulo, Brazil \\
\end{instit}

\medskip

\author{O. J. P. \'Eboli\cite{eb1} }

\smallskip

\begin{instit}
International Centre for Theoretical Physics, Trieste, Italy
\end{instit}

\begin{abstract}
We study the signals for composite vector leptoquarks in $e^+ e^-$
colliders (LEP II, NLC, and CLIC) through their effects on the
production of jet pairs, as well as their single and pair productions.
We also analyze their production in $\gamma e$ and $\gamma\gamma$
collisions.
\end{abstract}

\bigskip

\begin{center}
Submitted to Phys.\ Rev. D
\end{center}

\today

\newpage

\section{Introduction}

The standard electroweak theory provides a very satisfactory
description of most elementary particle phenomena up to the presently
available energies.  However, there are experimental facts such as the
proliferation of the fermion generations and their complex
pattern of masses and mixing angles, that are not predicted
by the standard model. A rather natural explanation for
the existence of the fermion generations is that the known particles
(leptons, quarks, and vector bosons) are composite. In general,
composite models exhibit a very rich spectrum which includes many new
states such as excitations of the known particles and bound states
which cannot be viewed as excitations of the familiar particles, since
they possess rather unusual quantum numbers.  Among these, there are
leptoquarks, which are particles carrying simultaneously leptonic and
baryonic number.  Leptoquarks are naturally present in a variety of
theories beyond the standard model such as some technicolor models
\cite{tec}, grand unified theories \cite{gut}, $E_6$
superstring-inspired models \cite{e6}, and composite models \cite{af}.

In the present work we study the production of vector leptoquarks in
$e^+e^-$, $e\gamma$, and $\gamma\gamma$ collisions. We shall consider
two sources of photons: they can be produced either by bremsstrahlung
or by backscattering laser light of the incident positron (electron)
beam \cite{laser}.  Here an intense hard-photon beam is generated by
backward Compton scattering of soft photons from a laser of a few eV
energy. We shall not consider beamstrahlung photons since its spectrum
depends strongly on the machine design.

For definiteness we shall consider the vector leptoquarks predicted by
the Abbott--Farhi model \cite{af}.  The Lagrangian of this model has
the same form as the standard model one.  However, the parameters
determining the potential for the scalar field and the strength of the
$SU(2)_L$ gauge interaction, are such that no spontaneous symmetry
breaking occurs and the $SU(2)_L$ gauge interaction is confining. The
model is essentially the confining version of the standard model and
is also called the strongly coupled standard model (SCSM). The
spectrum of physical particles in the SCSM consists of $SU(2)_L$ gauge
singlets, including fundamental particles which are neutral with
respect to the $SU(2)_L$ force, such as the right-handed fermions and
the $U(1)$ gauge boson.  For instance, the physical left-handed
fermions are bound states of a preonic scalar and a preonic dynamical
left-handed fermion, while the vector bosons are P-wave bound states
of the scalar preons.  Provided some dynamical assumptions on the
model hold true, it has been shown \cite{fj} that the predictions of
the SCSM model are consistent with the present experimental data.

We denote the preonic left-handed fermionic doublet by $\psi^a_L$,
with the flavor index $a$ running from 1 to 12 for three families.
$\psi^a_L$ belongs to a $\underline{2}$ representation of the
$SU(2)_L$ and to the $(0,\frac{1}{2})$ representation of the Lorentz
group.  The vector leptoquarks in the SCSM model are bound states of
the form $ {\psi^a_L}^\dagger \psi^b_L$, where $\psi^a_L$ carries
baryon number while $\psi^b_L$ carries lepton number. We define
$V_\mu^{ab}$ as the interpolating field for the vector leptoquarks,
which is an $SU(2)_L$ singlet, belongs to the
$(\frac{1}{2},\frac{1}{2})$ representation of the Lorentz group and is
a triplet under $SU(3)_{\rm color}$.  From its preonic content it
follows that these particles have an electric charge $-2/3$.

The SCSM model cannot be analyzed perturbatively since it is strongly
interacting at the energy scale of interest. Instead,  we describe the
interaction between leptoquarks and physical left-handed fermions by an
effective Lagrangian \cite{wud}.  We assume that

\begin{equation}
{\cal L}_{\rm int} = -F{e\over 2\sqrt{2}\sin^2\theta_W}
\left( {V_\mu^{ab}}^\dagger
\bar L^{a} \gamma^\mu  L^b + \mbox{h.c.} \right )
\label{lint}
\end{equation}
describes the low-energy interactions of $V_\mu^{ab}$, where $L^a$ are
physical left-handed doublets under the global $SU(2)$ symmetry of the
model \cite{fj}, and $\theta_W$ is the weak mixing angle.  The
parameter $F$ is a measure of the strength of this interaction
compared to the $Wq\bar q^\prime$ vertex. Notice that the vector
leptoquarks couple to both upper (or lower) components of the lepton
and quark doublets. It is important to realize that the above ${\cal
L}_{\rm int}$, conserves charge, color, and baryonic and leptonic numbers.

It is also natural to assume that vector leptoquarks $V_\mu^{ab}$
interact with the photon and the physical $Z$. In this work we assume
that the couplings of vector leptoquarks to $Z$'s and $\gamma$'s are
similar to the $W$ boson ones to these particles. Therefore, we
postulate the following Feynman rules (see Fig. (\ref{feyn:ru}))

\begin{equation}
\Gamma_{\alpha\beta\rho}^{\gamma V^-V^+} = ieQ_V \left \{ g_{\alpha\beta}
(p_1-p_2)_\rho + g_{\beta\rho} (p_2-p_3)_\alpha + g_{\rho\alpha}
(p_3-p_1)_\beta \right \} \; ,
\label{fvv}
\end{equation}

\begin{equation}
\Gamma_{\alpha\beta\rho\sigma}^{\gamma\gamma V^+V^-} = -i e^2Q_V^2 \left \{
2 g_{\alpha\beta} g_{\rho\sigma} - g_{\alpha\sigma} g_{\beta\rho}
- g_{\alpha\rho} g_{\beta\sigma} \right \} \; ,
\label{ffvv}
\end{equation}

\begin{equation}
\Gamma_{\alpha\beta\rho}^{Z V^-V^+} = - i F_Z^2 e \cot  \theta_W
 \left \{ g_{\alpha\beta}
(p_1-p_2)_\rho + g_{\beta\rho} (p_2-p_3)_\alpha + g_{\rho\alpha}
(p_3-p_1)_\beta \right \} \; ,
\label{zvv}
\end{equation}
where $Q_V$ ($=-2/3$) is the electric charge of the vector leptoquark
and $F_Z$ is a free parameter. The couplings in Eqs.
(\ref{fvv},\ref{ffvv}) were obtained via minimal substitution and
assuming that $V^{ab}_\mu$ has an anomalous magnetic moment
$\kappa=1$.

The absence of experimental evidence for compositeness constrains the low
energy phenomenology of the SCSM. These constraints can, in principle, place
bounds on the vector leptoquark mass ($M_V$) and coupling constants ($F$ and
$F_Z$). In fact, the analyzes of the contribution of vector leptoquarks to the
four-fermion Fermi interaction at low energies lead to the constraint
\cite{wud}

\begin{equation}
M_V > 197~ F ~(\hbox{\rm GeV}) \; .
\label{bound}
\end{equation}
In practice, contributions from other states soften this bound
\cite{kor}, so that $M_V$ and $F$ are in reality free parameters. However,
an educated guess for the coupling $F$ can be made as follows.  In the SCSM
model the $Z$ and the $W$ are bound states of two preonic scalars, therefore it
is natural to assume that the coupling of vector leptoquarks to physical
left-handed fields is of the same order of the coupling of these fermions to
$W$'s and $Z$'s, {\em i.e.\/}  $F$ is of order 1. Analogously, we expect that
$F_Z \simeq F \simeq O(1)$.

We can constrain the couplings $F$ and $F_Z$ imposing that unitarity
is respected at tree level \cite{corn}. For instance, the process $e^+
e^- \rightarrow V^+V^-$ violates unitarity at high energies for
arbitrary values of the couplings.  However, if we choose $F=F_Z=
\sqrt{|Q_V|}=\sqrt{\frac{2}{3}}$, unitarity at tree level is
restaured.

The main decay mode of vector leptoquarks are into a pair $lq$ or $\nu
q^\prime$, therefore its signal is a lepton plus a jet, or a jet plus missing
energy. Using the couplings given above we obtain that the width of a vector
leptoquark is given by

\begin{equation}
\Gamma_V = \frac{\alpha F^2}{4\sin^2\theta_W} M_V \; ,
\end{equation}
where we neglected all the fermion masses and summed over the possible
decay channels.

The outline of this paper is the following. The analisis of the
indirect signals for leptoquarks is contained in Sec. II: One way to
look for vector leptoquarks in $e^+e^-$ colliders is through their
effects on the production of jet pair ($e^+e^- \rightarrow q \bar q$),
since they can be exchanged in the $t$ channel. Another way tho search
for these particles is to study the forward--backward asymmetry in the
production of $b \bar b$ pairs.  In Sec. III, we study the single
production of vector leptoquarks through $e\gamma \rightarrow V^{eq}
q$, where the photons comes either from bremsstrahlung or from laser
backscattering.  In this Sec. we also discuss the signal and its
potential backgrounds.  Pairs of vector leptoquarks can also be
produced provided that there is enough available energy. Sec. IV
exhibits the study of the production of vector-leptoquark pairs in
$e^+e^-$ and $\gamma \gamma$ colliders.  We summarize our results on
Sec. V. The Appendix presents the relevant expressions for the photon
distribution functions used throughout this paper.

%^^^^^^^^^^^^^^^^^^^^^^^^^^^^^^^^^^^^^^^^^^^^^^^^^^^^^^^^^^^^^^^^^^^^^^^^^^^^^

\section{Indirect Evidence For Vector Leptoquarks}

We can look for signals of leptoquarks even when the available center
of mass energy is not enough to produce these particles on shell. This
can be done through the study of their effects as an intermediate
state of reactions like $e^+e^- \rightarrow $ dijets and $e^+e^-
\rightarrow b \bar b$.

\subsection{Total cross section $e^+e^- \rightarrow q \bar q$}

The existence of vector leptoquarks can be investigated through the
analyzes of the reaction $e^+e^- \rightarrow q \bar q$, where they
lead to a new $t$ channel contribution, in addition to the usual
exchange of $\gamma$ and $Z$ in the $s$ channel. Using the vertices
derived from the interaction Lagrangian (\ref{lint}), the cross
section for this process is given by

\begin{eqnarray}
\frac{d\sigma}{d\Omega} = && \frac{\alpha^2_{\rm em}}{4s} \Bigl \{
Q^2 (1 + \cos^2\theta) % \nonumber \\ &&
 + \frac{1}{16\sin^4\theta_W\cos^4\theta_W}
{}~ \frac{s^2}{(s-M_Z^2)^2+\Gamma_Z^2 M_Z^2}  \nonumber \\
&& \times \left [ \left ({C_V^e}^2 + {C_A^e}^2 \right ) \left (
{C_V^q}^2 + {C_A^q}^2
\right ) (1+\cos^2\theta) +8 C_V^e C_A^e C_V^q C_A^q \cos\theta \right ]
\nonumber \\
&& - \frac{Q}{2 \sin^2\theta_W \cos^2\theta_W }~\frac{s(s-M_Z^2)}
{(s-M_Z^2)^2+\Gamma_Z^2 M_Z^2} \left [
C_V^eC_V^q (1+\cos^2\theta) + 2 C_A^eC_A^q \cos\theta \right ]
\nonumber \\
&&+ \frac{F^2}{\sin^2\theta_W}~ \frac{(1+\cos\theta)^2}{\cos\theta-\eta}
\Bigl [ \frac{F^2}{4\sin^2\theta_W}~\frac{1}{\cos\theta-\eta}
+ \frac{Q}{2} \\
&&- \frac{1}{8\sin^2\theta_W\cos^2\theta_W}~ (C_V^q+C_A^q)(C_V^e+C_A^e)
{}~\frac{s(s-M_Z^2)}
{(s-M_Z^2)^2+\Gamma_Z^2 M_Z^2} \Bigr ] \Bigr \} \; ,
\nonumber
\end{eqnarray}
where $M_Z$ is the mass of the $Z$ boson, $\theta_W$ is the weak
mixing angle, and $\eta = 1 +2M_V^2/s$. According to our conventions
the charge of a quark is $Qe$ ($e> 0$), $ C_V=I_z-2Q\sin^2\theta_W$,
and $C_A = I_z$.

The exchange of a vector particle in the $t$ channel modifies the high
energy behaviour of this process: within the scope of the standard
model this cross section decreases as the center of mass energy
increases, however, the new contribution alters this behaviour,
yielding a constant cross section at high energies which is given by

\begin{equation}
 \sigma_{\rm limit} (e^+e^- \rightarrow q \bar q) \simeq \frac{\pi}{4}
\frac{\alpha^2 F^4}{\sin^4\theta_W}\frac{1}{M_V^2} \; .
\end{equation}
This is a dramatic signal once there will be many more dijets than the
expected in the scope of the standard model at high energies.  In Fig.
(\ref{sig:qq}), we exhibit the cross section $\sigma(e^+e^-
\rightarrow q \bar q)$ as a function of center of mass energy for
different values of the vector leptoquark mass and for $F=\sqrt{2/3}$.
This figure was obtained imposing the cut $|\cos \theta | < 0.9$, and
assuming the existence of three vector leptoquarks ($V^{ed}$,
$V^{es}$, $V^{eb}$), which have the same mass and values for the
coupling constants. Notice that, after the $Z$ peak the results, which
include the leptoquark, depart significantly from the standard model
prediction.

In order to estimate the capabilities of the different colliders (LEP
II, NLC, CLIC) to search for leptoquarks, we evaluate the largest mass
of a vector leptoquark, keeping $F$ fixed, for which the cross section
for dijet production differs by 10\% from the standard model result. In
our estimates we were conservative assuming that only one vector
leptoquark contributes to this reaction. We have defined

\begin{equation}
\Delta \equiv \frac{\sigma - \sigma_{WS}}{\sigma_{WS}} \; ,
\label{del}
\end{equation}
where $\sigma$ is the total cross section including the leptoquark
contribution and $\sigma_{WS}$ is the standard model result. Fig.
(\ref{del:qq}) displays $F$ as a function of $M_V$, which satisfies the
constraint $\Delta(F,M_V) = 10\%$, for several collider center of mass
energies.  From this figure, we can learn that an $e^+e^-$ collider
with center of mass energy of 200 (1000) GeV will be able to unravel
the existence of vector leptoquarks of masses up to 400 (2000) GeV,
assuming $F=\sqrt{2/3}$.

\subsection{Forward-backward asymmetry for $b \bar b$ pairs}

Another indirect way to look for the vector leptoquark $V^{eb}$ is
studying the forward-backward asymmetry in the production of $b \bar
b$ pairs.  Recently at LEP, this asymmetry has been measured
\cite{l3}, and it is in agreement with the standard model prediction.
Imposing that the contribution of this vector leptoquark to this
reaction is at most of the size of the experimental error (5\%), we
can exclude a region of the plane $M_V\times F$, as it is shown by the
dotted line in Fig. (\ref{afb:bb}).  Assuming $F=\sqrt{\frac{2}{3}}$,
the data constrains the mass of the $eb$ leptoquark to be bigger than
$\simeq 370$ GeV.

{}From Fig. (\ref{afb:bb}), we can also foresee the potential of the
future $e^+e^-$ machines for discovering the leptoquark $V^{eb}$: The
dashed, solid, and dot-dashed lines indicate the region for which the
forward-backward asymmetry is $5\%$, for center of mass energies of
$200$, $500$, and $1000$ GeV respectively.  For $F =
\sqrt{\frac{2}{3}}$, LEP II (NLC, CLIC) should be able to look for
$V^{eb}$ with mass up to $600$ ($1300$, $2300$) GeV.

%^^^^^^^^^^^^^^^^^^^^^^^^^^^^^^^^^^^^^^^^^^^^^^^^^^^^^^^^^^^^^^^^^^^^^^^^^^^^^^

\section{Single Production of Vector Leptoquarks}

We can produce a single vector leptoquark $V^{eq}$ ($q=d$, $s$, $b$)
through the process $ \gamma e^- \rightarrow V^{eq} q$. This process
can take place in $e^+e^-$ colliders, with the $\gamma$ being produced
by bremsstrahlung, or in $\gamma e$ machines, with the $\gamma$
originating from laser backscattering. The elementary cross section
for this reaction is

\begin{eqnarray}
\frac{d \hat \sigma}{d \hat t} &=& - N_c \frac{\pi}{36} \frac{F^2 \alpha^2}
{\sin^2\theta_W} \frac{ \left [ \hat s + 3  (\hat t - M_q^2) \right ]^2 }
{ M_V^2  (\hat s +\hat t - M_q^2)^2 (\hat t - M_q^2 )^2 \hat s^3}
\nonumber \\
&& \Bigl \{ (\hat t - M_q^2) \left [ M_q^2 (\hat s + \hat t)^2 + 2 \hat
s^2M_V^2  + 4 M_V^6 + M_q^6 \right ] \nonumber \\
&& - 4 \hat t M_V^4 (\hat s + \hat t) + 2 \hat t M_V^2 M_q^2 \left ( \hat s - 2
\hat t + M_V^2 + M_q^2 ) \right ] \\
&& + 2\hat s M_q^6 + 2 M_q^4 M_V^4 + 2 \hat t^3 M_V^2 \Bigl \} \; , \nonumber
\end{eqnarray}
where $N_c= 3 $ is the numbers of colors, $\hat s$ is the center of mass energy
squared of the subprocess, $\hat t = M_V^2 - \frac{\hat s}{2} (1 -\beta
\cos\theta^*)$, with $\beta$ being the $V^{eq}$ velocity in the subprocess c.m.
and $\theta^*$ its angle with respect to the incident electron in this frame.
In order to obtain the cross section for this reaction we must fold the above
expression with the $\gamma$ distribution function ($f_{\gamma/e}(x)$) (see
Appendix)

\begin{equation}
\sigma = \int_{x_{min}}^1 dx~ f_{\gamma/e}(x) \hat \sigma(xs) \; ,
\label{conv}
\end{equation}
where $x_{min} = (M_q+M_V)^2/s$.  Fig. (\ref{sig:eg}) exhibits the
behaviour of $\sigma$ as a function of $M_V$.  As expected, the
process initiated by laser backscattering possess a cross section that
is one order of magnitude larger than the processes initiated by
bremsstrahlung photons, with the same $M_V$ and $s$.

Once the leptoquark couples to $eq$ and $\nu q^\prime$ with the same strength,
the signal for its single production is either $(e)jj{\rm p}\hspace{-0.53
em}\raisebox{-0.27 ex}{/}_T$ or $(e)jje$, where the spectator $e$ is usually
lost in the beam pipe in the case of $e^+e^-$ colliders. The main background
for the signal $(e)jj{\rm p}\hspace{-0.53 em}\raisebox{-0.27 ex}{/}_T$
($(e)jje$) comes from the process $\gamma e \rightarrow W \nu$ ($e\gamma
\rightarrow Z e$) with the $W$ ($Z$) decaying into two jets \cite{eran}.
However, this background can be easily eliminated by requiring that the
invariant mass of the jet pair is not close to $M_W$ ($M_Z$).

At first sight, another potential background is the Bethe-Heitler
production of hadrons ($ \gamma e \rightarrow q \bar q$), which
exhibits a large cross section. However, the main contribution to the
cross section in this case is due to the region of small transverse
momenta of the produced particles. This allow us to reject with a high
efficiency this class of events by demanding that the observed
particles and jets have a sufficiently high $p_T$.

In order to access the capability of the future colliders to establish
the existence of leptoquarks through the reaction $e \gamma
\rightarrow V^{eq} q$, we require the occurance of $5000$ events
per year with the final state $jje^-$. Once the couplings $V^{eq}eq$
and $V^{eq}\nu q^\prime$ are expected to be approximately equal, we
take that $\sigma( jje^-) = \sigma(V^{eq}q) /2$. Assuming an integrated
luminosity of $10^{34}$ cm${}^{-2}$ s$^{-1}$ for the future machines,
the maximum observable mass for an $e^+e^-$ collider is $M_V= 300$
($400$) GeV for $\sqrt{s} = 500$ ($1000$) GeV, while a collider $\gamma e$
using laser backscattering can unravel the existence of leptoquarks of
mass up to $M_V=450$ ($900$) GeV for a center of mass energy of $500$
($1000$) GeV.

%^^^^^^^^^^^^^^^^^^^^^^^^^^^^^^^^^^^^^^^^^^^^^^^^^^^^^^^^^^^^^^^^^^^^^^^^^^^^^^

\section{Pair Production of Vector Leptoquarks}

\subsection{$e^+e^- \rightarrow V^+ V^-$}

Pairs of vector leptoquarks can be produced in $e^+e^-$ collisions provided
that there is enough available energy ($\sqrt{s} \ge 2 M_V$).  This process
takes place through the exchange of a quark in the $t$ channel and through
a $Z$ and $\gamma$ in the $s$ channel. Using the interaction Lagrangians of
Sect. I, it is easy to evaluate the cross section for this reaction, resulting
that

%\begin{flushleft}
\begin{eqnarray}
&&\frac{d\sigma}{dt} =
% canal t
\frac{F^4 \pi \alpha^2 }{16 s^2 t^2 M_V^4 \sin^4 \theta_W}
\left [ 3 s t^2 M_V^2 - 4 s M_V^6 + (t^2 + 4 M_V^4) (t-M_V^2) (u-M_V^2)
\right ] \nonumber \\
% gama no canal s
&& + \frac{\pi \alpha^2 Q_V^2}{2 s^4 M_V^4}
\Bigl [ t(s^2+tM_V^2)(u-M_V^2) + s^2 M_V^2 (s - u - 10 M_V^2) + 2 st^2M_V^2
+ stu M_V^2 \nonumber \\
&& - 4 s M_V^4 (t-u) + 2 su^2M_V^2 + tu^2M_V^2 + 8tuM_V^4
- u^2 M_V^4 - 8 M_V^8 \Bigr ]  \nonumber \\
% interferencia gama canal t
&& + \frac{F^2\pi\alpha^2 Q_V}{4s^3tM_V^4\sin^2\theta_W}
\Bigl [ s^2 t M_V^2 + 4 s^2 M_V^4 + s t^2 (u - 3 M_V^2) - 3 s t M_V^2
(u-M_V^2) + 4 s M_V^4 (u+M_V^2) \nonumber \\
&& - 2 t^2 M_V^2 (u-M_V^2) - 2 t M_V^4 (u-M_V^2) + 4 M_V^6 (u-M_V^2)
\Bigr ] \nonumber \\
% Z_0 no canal s
&& + \frac{F_Z^4\pi\alpha^2}{8s^2M_V^4\sin^4\theta_W} (C_V^2+C_A^2)
\frac{1}{( (s-M_Z^2)^2 + \Gamma_Z^2 M_Z^2) } \Bigl [ s^2 M_V^2 (s -t-u-10
M_V^2) + s^2 t u + 8 t u M_V^4 \nonumber \\
&& + s t M_V^2 (2t + u - 4 M_V^2) + 2 suM_V^2 (u - 2M_V^2) + t^2 M_V^2
(u-M_V^2)
+ u^2 M_V^2 (t-M_V^2)  - 8 M_V^8 \Bigr ]  \nonumber \\
% interferencia Z-gama
&& + \frac{F_Z^2 \pi \alpha^2 C_V Q_V}{2 s^3 M_V^4 \sin^2\theta_W}
\frac{s-M_Z^2}{((s-M_Z^2)^2 + \Gamma_Z^2 M_Z^2)}
\Bigl [ s^2 M_V^2 (s-u-10M_V^2) + s^2 t (u-M_V^2) +t^2M_V^2 (u-M_V^2)
\nonumber \\
&& + s t M_V^2 ( 2 t + u - 4 M_V^2) + 2suM_V^2 (u-2 M_V^2)
+u^2 M_V^2(t-M_V^2) + 8tuM_V^4   -8 M_V^8 \Bigr ] \nonumber \\
% interferencia Z canal t
&& + \frac{F_Z^2 F^2 \alpha^2}{8s^2tM_V^4\sin^4\theta_W}
(C_V+C_A) \frac{s-M_Z^2}{((s-M_Z^2)^2 +\Gamma_Z^2 M_Z^2)}
\Bigl [ s^2 t M_V^2 + 4 s^2 M_V^4 + s t^2 u \nonumber \\
&& - 3 s t M_V^2 (t + u -M_V^2) + 4s M_V^4 (u+M_V^2) - 2 t M_V^2 (u-M_V^2)
(t + M_V^2) + 4 M_V^6 (u-M_V^2) \Bigr ] \nonumber
\label{pair}
\end{eqnarray}
where $t=M_V^2- \frac{s}{2}(1-\beta\cos\theta)$, with $\theta$ being
the scattering angle between the $e^-$ and the negatively charged
leptoquark in the laboratory frame, and $\beta =
\sqrt{1-4M_V^2/s}$. This cross section exhibits a bad high energy
behaviour ($\sigma \propto s$), and violates unitarity in this limit
for an arbitrary choice of the couplings $F$ and $F_Z$.  However, this
violation of unitarity can be avoided by a careful choice of the
couplings: for $F=F_Z=\sqrt{\frac{2}{3}}$ this cross section has a
good high energy behaviour. Moreover, for these values of the
couplings, the cross section for this process is $4/9  \sigma(
e^+e^- \rightarrow W^+W^-)$. Therefore, we must make this
choice if we want to preserve unitary at tree level.

Fig. (\ref{sig:eevv}) exhibits the total cross section for the process $e^+e^-
\rightarrow V^+V^-$ as a function of $M_V$ for $F=F_Z=\sqrt{\frac{2}{3}}$. The
signal for such a process is either $jjee$, $jje{\rm p}\hspace{-0.53
em}\raisebox{-0.27 ex}{/}_T$, or $jj{\rm p}\hspace{-0.53 em}\raisebox{-0.27
ex}{/}_T$. Certainly the identification of the leptoquark is very easy in the
mode $jjee$ since the backgrounds, like $e^+e^- \rightarrow ZZ$, can be
efficiently eliminated by looking at the invariant mass of the pairs $ee$
and/or $jj$. Moreover, the signal is very striking since it consists of two
pairs $ej$ with (approximately) the same invariant mass.  Assuming an
integrated luminosity of $10^5$ pb${}^{-1}$ per year, there will be more than
$10^5$ events per year, which is more than enough to establish the existence of
the leptoquarks.

\subsection{$\gamma\gamma \rightarrow V^+ V^-$}

We can also produce pairs $V^+V^-$ in $\gamma\gamma$ collisions, where
the photons are generated either by bremsstrahlung or by laser
backscattering.  There are three Feynman diagrams that contribute to
this process: there is the exchange of a $V$ in the $t$ and $u$
channels and the quartic vertex $\gamma\gamma VV$.  The cross section
for this reaction is equal to the one for $\gamma\gamma
\rightarrow W^+ W^-$ scaled by a factor $Q_V^4$, since the couplings
$V\gamma$ and $W\gamma$ are assumed to be proportional. It is
straightforward obtain that the subprocess cross section is

\begin{equation}
\frac{d\hat \sigma}{d\hat t} = Q_V^4 \frac{8\pi \alpha^2}{M_V^2}
\left [  \frac{(16x^2+3)M_V^6}{2(\hat t - M_V^2)^2 (\hat u - M_V^2)^2}
- \frac{(8x+3)M_V^2}{8x(\hat t - M_V^2)(\hat u-M_V^2)} +
\frac{3}{64 x^2 M_V^2} \right ] \; ,
\label{sig:ggvv}
\end{equation}
where we defined $x = \hat s / 4 M_V^2$.  One characteristic of this
process is that the cross section is peaked at the forward region at
high energies. Furthermore, due to exchange of a spin-$1$ particle in the
$t$ and $u$ channels, the cross section goes to a constant at high energies

\begin{equation}
\hat \sigma_{\rm limit} \simeq \frac{128}{81} \frac{\pi\alpha^2}{M_V^2} \; .
\end{equation}

We can obtain the total cross section for this process folding $\hat \sigma$
with the photon distribution functions.

\begin{equation}
\sigma = \int dx_1 \int dx_2 f_{\gamma/e}(x_1) f_{\gamma/e}(x_2)
\hat \sigma (\hat s = x_1 x_2 s)
\end{equation}

Figs. (\ref{sig:ggvv_br}) and (\ref{sig:ggvv_la}) show the behaviour
of the cross section of the process $\gamma\gamma \rightarrow V^+V^-$ as a
function of $M_V$ for bremsstrahlung and laser backscattering photons
respectively. In this case also, the cross section for the process
initiated by backscattered photons is one to two orders of magnitude
larger than the one for bremsstrahlung photons due to the distribution
of backscattered photons being harder than the one for bremsstrahlung.
{}From Fig. (\ref{sig:ggvv_br}) we can infer that this reaction is
observable for leptoquark masses up to $\simeq 100$ ($200$) GeV in an
$e^+e^-$ machine with $\sqrt{s} = 500$ ($1000$) GeV. Analogously, we
can see from Fig. (\ref{sig:ggvv_la}), that this process is observable
for masses up to $\simeq 200$ ($400$) GeV in a $\gamma\gamma$ collider
with $\sqrt{s} = 500$ ($1000$) GeV.

\section{Conclusions}

We studied the signals of vector leptoquarks in $e^+e^-$, $e\gamma$,
and $\gamma\gamma$ machines. In order to do so, we postulated the
interaction Lagrangian of the vector leptoquarks with the quarks and
leptons. Demanding that unitarity is satisfied at tree level, in the
different process analyzed \cite{corn}, we discovered that the
couplings $F$ and $F_Z$ are constrained to the value $\sqrt{2/3}$.

In $e^+e^-$ machines, we can look for these particles through their
effect in $e^+e^- \rightarrow q \bar q$, and the existence of
leptoquark can be established provided their masses are smaller than
$O(2\sqrt{s})$. In these machines, vector leptoquarks can also be single
produced through the reaction $e\gamma \rightarrow V^{eq} q$, where
the $\gamma$ originates from bremsstrahlung. This process is observable
for leptoquark masses up to $300$ ($400$) GeV, in a collider with
$\sqrt{s} = 500$ ($1000$) GeV. We have also studied the production of
leptoquark pairs, either through $e^+e^- \rightarrow V^+V^-$ or
$\gamma \gamma \rightarrow V^+V^-$, with the two photons coming from
bremsstrahlung.

In a $e \gamma$ collider, leptoquarks can be produced in association
with jets through $e\gamma \rightarrow V^{eq} q$. For hard photons
produced by laser backscattering, we can detect this process provided
that the leptoquark mass is smaller than $450$ ($900$) GeV, for a
collider with $\sqrt{s}=500$ ($1000$) GeV. We also investigated the
production of $V^+V^-$ pairs in $\gamma\gamma$ collisions, and we
found that this processes can be observed for leptoquarks with
masses up to $200$ ($400$) GeV, if $\sqrt{s} = 500$ ($1000$) GeV.

\nonum
\section{Acknowledgments}

This work was partially supported by Conselho Nacional de
Desenvolvimento Cient\'\i fico e Tecnol\'ogico (CNPq), Funda\c{c}\~ao
de Amparo \`a Pesquisa do Estado de S\~ao Paulo (FAPESP), and
Coordenadoria de Aperfei\c{c}oamento de Pessoal de N\'\i vel Superior
(CAPES). One of the authors (OJPE) would like to thank the hospitality
of the ICTP-Trieste, where part of this work was carried on.  We also
would like to thank prof. S.F.\ Novaes for a careful reading of the
manuscript.

\unletteredappendix{}

%^^^^^^^^^^^^^^^^^^^^^^^^^^^^^^^^^^^^^^^^^^^^^^^^^^^^^^^^^^^^^^^^^^^^^^^^^^^^^

The contribution arising from the conventional bremsstrahlung photons
were computed using the well-known Weisz\"acker-Willians distribution
\cite{brem}

\begin{equation}
f^{ww}_{\gamma/e}(x) =  \frac{\alpha}{2\pi} \frac{1+(1-x)^2}{x} \ln
\left ( \frac{s}{4m_e^2} \right ) \; ,
\end{equation}
where $m_e$ is the electron mass, and $s$ is the center of mass energy of the
$e^+e^-$ pair. This spectrum is peaked at small $x$, {\em i.e.\/} most of its
photons are soft.

Hard photons can be obtained by laser backscattering, which converts an $e$
beam into a $\gamma$ one. Here the intense photon beams is generated by
backward Compton scattering of soft photons from a laser of a few eV energy.
The energy spectrum of the backscattered laser photons is \cite{laser}

\begin{equation}
f_{\gamma/e}^L (x,\xi) \equiv \frac{1}{\sigma_c} \frac{d\sigma_c}{dx} =
\frac{1}{D(\xi)} \left[ 1 - x + \frac{1}{1-x} - \frac{4x}{\xi (1-x)} +
\frac{4
x^2}{\xi^2 (1-x)^2}  \right] \; ,
\label{F:laser}
\end{equation}
where $\sigma_c$ is the total Compton cross section. For the photons going in
the direction of the initial electron, the fraction $x$ represents the ratio
between the scattered photon and the initial electron energy ($x = \omega/E$).
In writing Eq. (\ref{F:laser}), we defined

\begin{equation}
D(\xi) = \left(1 - \frac{4}{\xi} - \frac{8}{\xi^2}  \right) \ln (1 + \xi) +
\frac{1}{2} + \frac{8}{\xi} - \frac{1}{2(1 + \xi)^2} \; ,
\label{D}
\end{equation}
with

\begin{equation}
\xi \equiv \frac{4 E \omega_0}{m^2} \cos^2 \frac{\alpha_0}{2} \simeq
\frac{2 \sqrt{s} \omega_0}{m^2} \; ,
\label{xi}
\end{equation}
where $\omega_0$ is the laser photon energy and ($\alpha_0 \sim 0$) is the
electron-laser collision angle. It is easy to verify that
the maximum value of $x$ possible in this process is
\begin{equation}
x_m = \frac{\omega_m}{E} =  \frac{\xi}{1 + \xi} \; .
\label{ym}
\end{equation}
{}From Eq. (\ref{F:laser}) we can see that the fraction of photons with
energy close to the maximum value grows with $E$ and $\omega_0$.
Usually, the choice of $\omega_0$ is such that it is not possible for
the backscattered photon to interact with the laser and create
$e^+e^-$ pairs, otherwise the conversion of electrons to photons would
be dramatically reduced. In our numerical calculations, we assumed
$\omega_0 \simeq 1.26$ eV, which is below the threshold of $e^+e^-$
pair creation ($\omega_m \omega_0 < m^2$). Thus for the NLC beams
($\sqrt{s} = 500$ GeV), we have $\xi \simeq 4.8$, $D(\xi) \simeq 1.9$,
and $x_m \simeq 0.83$.

%************** Figuras

\figure{Feynman rules for the vertices $\gamma V^+V^-$, $\gamma\gamma V^+V^-$,
and $Z_0V^+V^-$. \label{feyn:ru} }

\figure{Total cross section for the production of two jets as a function of the
collider center of mass energy. The solid line stands for the standard
model result, while the dotted, dashed, and dot-dashed lines include
the contribution of a vector leptoquark of mass 300, 700, and 1500 GeV
respectively. \label{sig:qq} }
%OK

\figure{$F$ as a function of $M_V$ for several $\sqrt{s}$: the dotted, dashed,
solid, and dot-dashed lines stand for $\sqrt{s}=100$, $200$, $500$,
and $1000$ GeV respectively.  \label{del:qq} }
%OK

\figure{Allowed values of the coupling $F$ and $M_V$ from the experimental
results from LEP for $b \bar b$ production (dotted line).  The dashed,
solid, and dot-dashed lines are the region for which the
forward-backward asymmetry is $5\%$, for center of mass energies of
$200$, $500$, and $1000$ GeV respectively.
\label{afb:bb} }
%OK

\figure{ Total cross section for the process $e^- \gamma \rightarrow
V^- q$ as a function of $M_V$: (a) laser backscattering at $\sqrt{s} =
500$ GeV (dotted line); (b) laser backscattering at $\sqrt{s} = 1000$
GeV (solid line); (c) bremsstrahlung at $\sqrt{s} = 500$ GeV
(dot-dashed line); (d) bremsstrahlung at $\sqrt{s} = 1000$ GeV (dashed
line). \label{sig:eg} }
%OK

\figure{Cross section of the process $e^+ e^- \rightarrow V^+ V^-$
as a function of $M_V$ for several collider energies: $\sqrt{s}= 500$
(dotted line); $1000$ (solid line); $2000$ (dashed line) GeV. It
was assumed that $F=F_Z=\sqrt{\frac{2}{3}}$. \label{sig:eevv} }
%OK

\figure{Cross section for the process $\gamma \gamma \rightarrow
V^+V^-$, with the $\gamma$ originating from bremsstrahlung, for
several energies: (a) $\sqrt{s}= 500$ (dotted line); (b) $1000$ (solid line);
(c) $2000$ GeV (dot-dashed line). \label{sig:ggvv_br} }
%OK

\figure{Same as in Fig. (\ref{sig:ggvv_br}), but with the $\gamma$'s
produced by laser backscattering. \label{sig:ggvv_la} }
%OK

\end{document}